\documentclass[preprint,12pt]{elsarticle}

\usepackage{amssymb}
\usepackage[normalem]{ulem}

\newcounter{bla}

\emergencystretch=\maxdimen
\journal{Computer Physics Communications}

\begin{document}

\begin{frontmatter}


\title{PARCE: Protocol for Amino acid Refinement through Computational Evolution}

\author[a]{Rodrigo Ochoa}
\author[b]{Miguel A. Soler}
\author[c,d]{Alessandro Laio}
\author[a,e]{Pilar Cossio\corref{author}}

\address[a]{Biophysics of Tropical Diseases, Max Planck Tandem Group, University of Antioquia, 050010 Medellin, Colombia}
\address[b]{Italian Institute of Technology (IIT), Via Melen 83, B Block, 16152, Genova, Italy}
\address[c]{International School for Advanced Studies (SISSA), Via Bonomea 265, I-34136 Trieste, Italy}
\address[d]{The Abdus Salam International Centre for Theoretical Physics (ICTP), Strada Costiera 11, 34151 Trieste, Italy}
\address[e]{Department of Theoretical Biophysics, Max Planck Institute of Biophysics, 60438 Frankfurt am Main, Germany}

\cortext[author] {Corresponding author.\\\textit{E-mail address:} grupotandem.biotd@udea.edu.co; picossio@biophys.mpg.de}

\begin{abstract}
The \textit{in silico} design of peptides and proteins as binders is useful for diagnosis and therapeutics due to their low adverse effects and major specificity. To select the most promising candidates, a key matter is to understand their interactions with protein targets. In this work, we present PARCE, an open source Protocol for Amino acid Refinement through Computational Evolution that implements an advanced and promising method for the design of peptides and proteins. The protocol performs a random mutation in the binder sequence, then samples the bound conformations using molecular dynamics simulations, and evaluates the protein-protein interactions from multiple scoring. Finally, it accepts or rejects the mutation by applying a consensus criterion based on binding scores. The procedure is iterated with the aim to explore efficiently novel sequences with potential better affinities toward their targets. We also provide a tutorial for running and reproducing the methodology.

\end{abstract}

\begin{keyword}
Amino acids; Bioinformatics; Molecular simulations; Mutation

\end{keyword}

\end{frontmatter}

% \linenumbers

% Computer program descriptions should contain the following
% PROGRAM SUMMARY.

{\bf PROGRAM SUMMARY}
  %Delete as appropriate.

\begin{small}
\noindent

{\em Program Title: PARCE}                                          \\
{\em Licensing provisions: MIT License}                                   \\
  %enter "none" if CPC non-profit use license is sufficient.
{\em Programming language: Python 3}                                   \\
{\em Operating system:  Linux, and docker container available to run the protocol under other operating systems.}                                       \\
  %Operating system(s) for which program has been designed.
{\em Subprograms used: Gromacs 5.1.4, Scwrl 4, PDB2PQR, Scoring functions source code.}                                       \\
  %Fill in if necessary, otherwise leave out.
{\em Nature of problem: Computational design of peptides and proteins as binders for diagnosis and therapeutics.}
  %Describe the nature of the problem here.
   \\
{\em Solution method:  A protocol that performs random mutations in the binder sequence, samples the bound conformations using molecular dynamics simulations, and evaluates the protein-protein interactions from multiple scoring predictions in order to accept or reject the mutations.}
  %Describe the method solution here.
   \\
{\em Additional comments: This article describes version 1.0. PARCE is available at: https://github.com/PARCE-project/PARCE-1, and a Docker container can be downloaded from: https://hub.docker.com/r/rochoa85/parce-1}\\
  %Provide any additional comments here.
   \\

\end{small}


\section{Introduction}

The rational design of proteins and peptides is a field that has been explored through experimental and computational approaches \citep{sormanni2018third}. When detailed information of a particular system is available, conventional knowledge-based methods provide tools for the design of more active analogs \citep{Juretic2011}. However, detailed information is not always available, and thus, \textit{de novo} design strategies are required.

Due to the significant amount of information related to the properties of natural amino acids, one strategy is to design proteins and peptides by modifying directly the amino acid sequence, without taking into account the structure. This strategy is commonly used in the design of antimicrobial peptides, where the mechanism of action has been elucidated for some families \citep{Porto2012}. Physico-chemical properties such as hydrophobicity, charge, and molecular weight, have been used to create libraries of analogs with potentially similar activities \citep{Boone2018}. However, these strategies lack information about the interacting partners, and on how the conformational landscape of the residues can affect their activity \citep{Ochoa2018}. This motivates the use of structure-based design \citep{Hansen2017,Guida2017}.

The most used structure-based strategies are molecular docking and the use of structural libraries. Approaches based on experimental structural databases take advantage on the huge quantity of protein structural information to predict and optimize the most probable binding conformations \citep{sormanni2015rational,van2018improved,adolf2018rosettaantibodydesign}. Molecular docking identifies the most probable poses of potential active molecules ranked by scoring functions that roughly approximate binding free energies or discriminate between active molecules and decoys \citep{Moal2013,Bohm1999}. In spite of these scoring-function improvements, protein and peptide design is still a computational challenge due to the large sequence space that has to be explored (\textit{e.g.}, for a peptide of $10$ amino acids in length, $20^{10}$ sequences are possible). In addition, the intrinsic flexibility of non-structured protein regions, such as peptides or the complementary determining regions of antibodies,  and the constraints associated to the protein interface region affect the reliability of the prediction \citep{Kurcinski2015}. To solve some of these problems, it has been proposed to sample the bound conformations using Molecular Dynamics (MD) \citep{Gladich2015}.

The prediction of the binding affinities is another challenge associated to the design. Different alternatives exist based on the biological system and the available computational infrastructure. One option is the use of representative structures from MD trajectories to calculate energy terms based on molecular mechanics assumptions, entropy terms and free energies of solvation using continuum-solvent models \citep{Genheden2015,Obiol-Pardo2007,Wichapong2016}. However, these approaches are computationally expensive. An alternative is averaging the scoring over conformational ensembles to estimate the affinity \citep{Soler2018,Ochoa2016,sarti2016protein}.

\begin{figure*}[hbt!]
\centering
\includegraphics[width=1.0\textwidth]{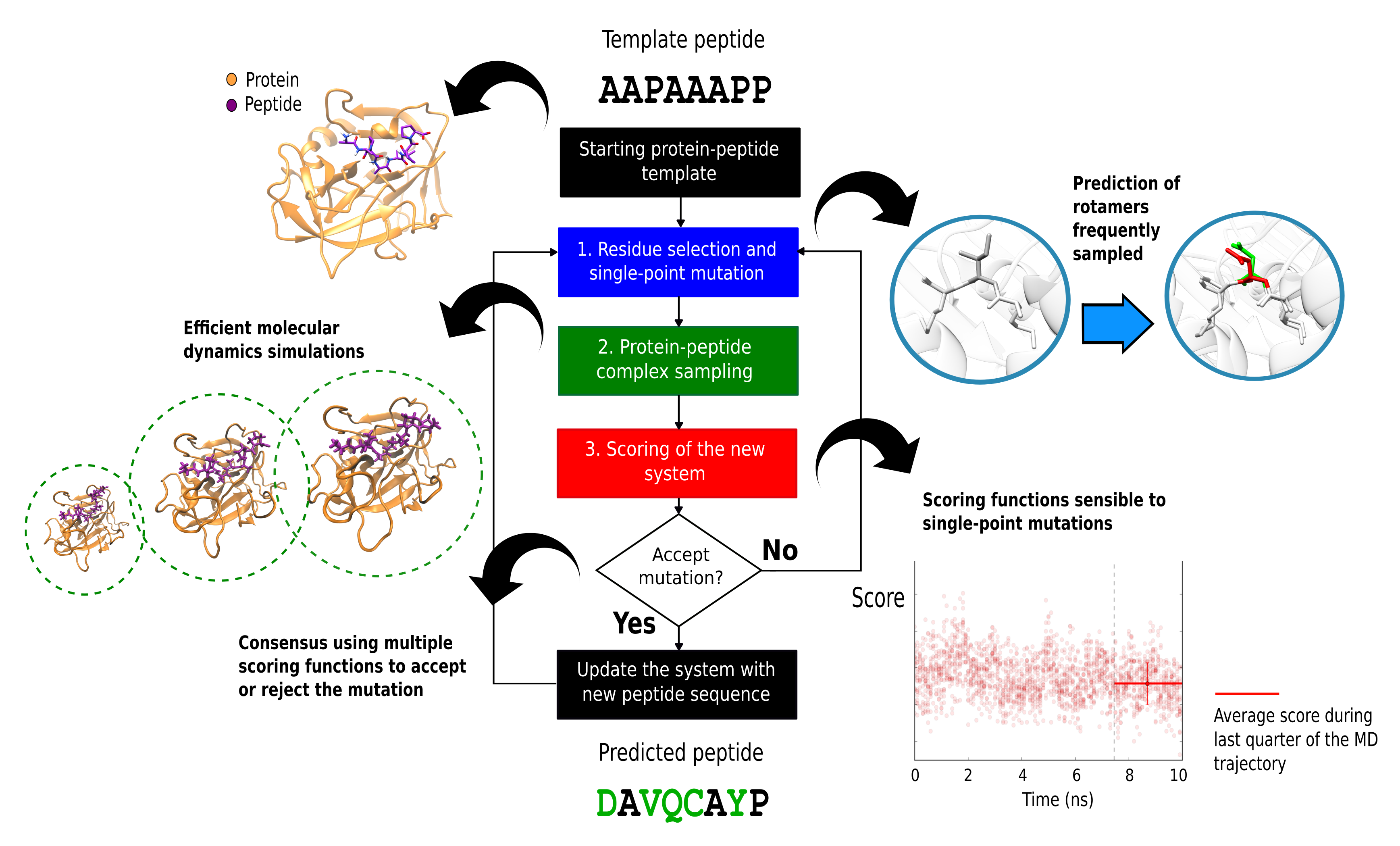}
\caption{Summary of PARCE, which is composed of different phases. First, a starting protein-peptide or protein-protein complex is used. Then, the MC cycle starts by generating a single-point mutation of a random amino acid in the peptide structure. Then, the conformational sampling of the mutated system is done with MD simulations. Scoring of the trajectory snapshots and the calculation of averages using multiple scoring functions is performed. Finally, a consensus criterion accepts or rejects the mutation and an update (or not) of the sequence is performed. The process is iterated.}
\label{fig:Fig1}
\end{figure*}

We here describe a new tool for performing peptide and protein design using MD. We called this tool Protocol for Amino acid Refinement through Computational Evolution (PARCE). The algorithm implemented in PARCE is inspired by previous projects related to the exploration of the sequence space of peptides bound to proteins or small molecules \citep{HongEnriquez2012,Gladich2015,Russo2015}. The algorithm includes a single-point mutation protocol, which has been optimized for predicting the most common rotamers of peptide amino acids \citep{Ochoa2018}, an all-atom MD simulation of the mutated complex and the estimate of the average score of the conformations from the trajectory to assess the impact of the mutation in the binding \citep{Gladich2015,Soler2017}. This approach has been successfully used to design peptides and protein fragments, whose binding affinity was afterward confirmed experimentally \citep{Ochoa2019,Soler2018,Soler2019}. PARCE implements multiple scoring functions that are evaluated using a consensus metric for accepting or rejecting the mutations \citep{Soler2019}. The process is iterated with the aim to evolve the original sequence and explore efficiently novel sequences with potential better affinities toward their targets \citep{Guida2017}. Figure \ref{fig:Fig1} shows a graphical summary of PARCE.

The computational protocol presented in this work is open source. The manuscript is organized as follows. First, we explain the code architecture, the simulation parameters and the system used to test the PARCE functionalities. Then, we include details about the code requirements and the performance of the protocol using a protein-peptide system as a reference. Finally we discuss PARCE features in comparison with those of similar software.

\section{Methods}

\subsection{Code architecture}
The code is written in Python 3, with calls to Unix command lines for manipulation of files and external programs. Therefore, the current protocol depends on the availability of a bash environment to run. To guarantee reproducibility, a docker container is provided with all the requirements, including the installation of the third-party open source software, and the validation of the PARCE functionalities. However, a user can configure her/his own machine following a simple setup validated with the Travis CI framework (https://travis-ci.com). The documentation illustrates the definition of the parameters, the input options and files (see supplementary Table \ref{tbl:suptable1}). Figure \ref{fig:Fig2} shows a workflow of the protocol based on the tasks, scripts and software implemented at each step. In the following, we present a description of each stage of PARCE.

\begin{figure*}[hbt!]
\centering
\includegraphics[width=1.0\textwidth]{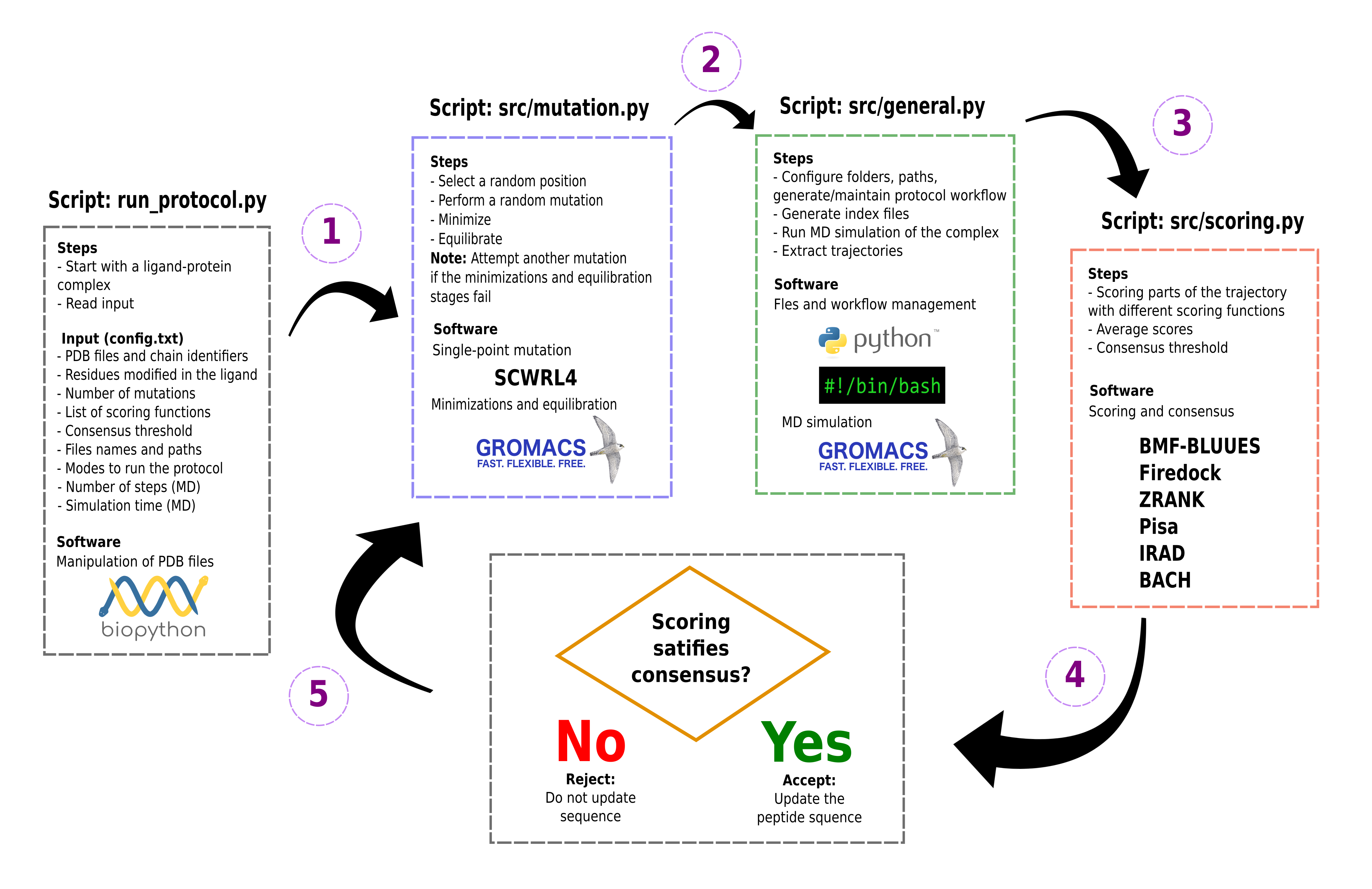}
\caption{Workflow of PARCE. The protocol is managed by the script \textit{run\_protocol.py}, which accepts as input a configuration file with all the parameters required to run the design, and a folder with the structural input obtained from a previous MD trajectory. After configuring all the files and folders, and runing MD simulations of the initial complex, the first step is to perform a single-point mutation with the \textit{mutation.py} script, which selects a random position on the peptide and modifies its side chain by a random amino acid. From there, the second step is to call the \textit{general.py} script, responsible to sample the new system. The third step is to score the trajectory using a set of scoring functions available in the \textit{scoring.py} file. The fourth step is to verify if the score difference between the current state and the previous one satisfies a consensus threshold. If that happens then the system is updated. The process is repeated during a number of attempts defined by the user.}
\label{fig:Fig2}
\end{figure*}

\subsubsection{Input requirements and initial MD simulation}
Before applying the design protocol it is necessary to perform an MD simulation of the starting complex to equilibrate the system, and also to obtain MD files required to start the protocol. When the starting system is obtained from a docked structural complex or a crystal structure, a longer previous sampling of the complex is suggested. The equilibration of the complex can be detected using the RMSD or observables such as the number of hydrogen bonds between the peptide and protein. We note that for most systems 200 ns is suitable for this purpose. If the protocol starts from a crystallized protein-peptide complex, the initial sampling time could be lower \citep{Ochoa2019}. 

The workflow requires as input:
\begin{itemize}
    \item A PDB and GRO (Gromacs format) file of the protein-peptide complex solvated and equilibrated with MD using the Gromacs package \citep{Hess2008}. 
    \item Topology files and MD parameters definitions.
\end{itemize}

The protocol starts with an initial NPT simulation of the system using the Gromacs configuration files (\textit{i.e.} mdp files). Advanced users have the possibility of modifying the Gromacs configuration files for adapting these to simulate the specific systems. 

\subsubsection{Main script}

The code has the main script \textit{run\_protocol.py}, which calls three main modules: general functions, scoring functions and mutation functions. The general module creates an object responsible to run the simulations, perform the random mutations and score the trajectories to accept or reject the mutations. In a folder called \textit{design\_output}, all the results are stored step-by-step, including the different molecular entities (in complex or isolated), as well as the MD trajectories, the log files and the scores calculated for each frame. In the end, a report file summarizes the mutations attempted, if accepted or rejected, the average scores of each function and the updated peptide sequence. An example is provided in supplementary Table \ref{tbl:suptable2}. The number of MC steps and the MD simulation time can be modified for making the protocol computationally efficient.

\subsubsection{Mutation protocol}
The script \textit{mutation.py} randomly performs a single-point mutation over the peptide. Currently, the code randomly selects a residue on the peptide chain and performs a random mutation using a uniform distribution for the aminoacids. However, this procedure can be customized, for example, by avoiding certain problematic amino acids, such cysteines, or including a non-uniform probability distribution for the location or aminoacid-type mutation.

The prediction of the new amino acid rotamer is made with Scwrl4 \citep{Krivov2009,Peterson2014}, which selects the side chain conformations based on a library of backbone-dependent rotamers previously extracted from crystallized protein structures. The program was chosen based on a previous assessment of different mutation protocols to predict amino acid rotamers most similar to those frequently explored by MD trajectories of protein-peptide systems \citep{Ochoa2018}. The mutation is first generated over the complex without solvent. Then a first minimization is ran with the predicted side chain alone. Then the mutated complex is combined with the equilibrated solvent box from the input structure file, and a second minimization is ran with the new amino acid and the water molecules surrounding it within 2\AA. This is done to avoid clashes between the new side chain and water molecules, which could crash the simulation. If the minimization crashes, the protocol attempts a new mutation, which is be repeated based on a number of times defined at the beginning by the user. If the protocol continues presenting minimization problems, the new mutation is generated using as reference the last accepted sequence. If the problems persists, the protocol stops. In addition, it is possible to minimize the input structure before mutating to avoid protein-solvent overlaps. Finally, the complete system is minimized and subjected to NVT equilibration of 100 picosencods (ps). Details of the MD simulations are described next.

\subsubsection{Conformational sampling}

The script \textit{general.py} runs the MD simulations using Gromacs \citep{Hess2008}. The current PARCE code was tested using version 5.1.4, but any higher version can be also selected by providing the path in the configuration file (see supplementary Table \ref{tbl:suptable1}). The MD simulation time is defined by the user, which by default is 10 nanoseconds (ns) per run. This production run is performed in the NPT ensemble after minimizing and equilibrating the system. The Amber99SB-ILDN protein force field \citep{Lindorff-Larsen2010} is selected, with a TIP3P water model \citep{Jorgensen1983}, a Parrinelo-Rahman barostat \citep{Parrinello1980} and a modified Berendsen thermostat \citep{bussi2007canonical}. The Particle Mesh Ewald method (PME) is used to calculate electrostatic interactions with a threshold of 1.0 nm \citep{DiPierro2015}. The leap-frog integrator \citep{Janezic1995} is used to integrate the equations of motion with a timestep of 2 femtoseconds (fs). A standard temperature of 310K is selected by default to run the simulations. We note that these variables can be modified directly in the Gromacs configuration files depending on the system and the preferred conditions. After each run, the trajectory and the simulation-log files are stored for further analysis.

\subsubsection{Scoring functions}
The script \textit{scoring.py} implements a scoring approach over 
the conformations from the complex trajectory. Its objective is to rank the mutated complex by comparing the predicted activity to that of the previous complex. A set of scoring functions used for protein-protein and protein-peptide affinity predictions are implemented to score each snapshot and obtain an average score for each function. The average can be calculated over all the trajectory frames, or just the last half of the simulation. The code includes six scoring functions: BACH \citep{Cossio2012,sarti2013bachscore,Sarti2015}, Pisa \citep{Krissinel2007}, FireDock \citep{Andrusier2007}, Irad \citep{Vreven2011}, Zrank \citep{Pierce2007}, BMF-Bluues \citep{Berrera2003,Fogolari2012} and DFIRE-GOAP \citep{Yang2008,Zhou2011} combinations. These software packages are open source and are distributed with the PARCE code.

\subsubsection{Consensus strategy}
The mutation is accepted or rejected based on a consensus criterion using multiple scoring functions. The sign of the score difference between the new sequence and the previous one indicates if the mutation is favorable or not for each scoring function. In this scenario, a negative sign of the difference means a potentially better affinity when the mutation is performed. The consensus criterion states that if the number of scoring functions that consider the mutation favorable is higher than a defined threshold, then mutation is accepted \citep{Soler2019}. By default the consensus threshold is three, but this value can be changed by the user. The implemented scores can be customized by the user, and additional scoring functions can be added if necessary. After that, the protocol iterates over the three main phases (mutating, sampling and scoring) and a number of attempted modifications is achieved (Figure \ref{fig:Fig2}).

\subsubsection{Output files}

After running PARCE with a system of reference, a set of folders containing structures of each iteration, the MD trajectories, the calculated scores and the sampling log files are generated locally. The explanation of the folders is provided in supplementary Table \ref{tbl:suptable3}. The files and final report can be used for further analysis of the best sequence candidates.

\section{Results}

\subsection{Code insights}

PARCE can be installed under any Linux operating system. However, the code was optimized for Debian and Ubuntu OS distributions. PARCE can be managed also through a docker container provided in the repository. We should note that the computational resources required for running PARCE are determined by the complexity of the system, since the design is based on running MD. A configuration file describes the path and characteristics of the input files, as well as the necessary parameters to run the design protocol. An explanation of the parameters and output files is provided in Supplementary tables. The code contains instructions to configure the system and launch the protocol. We note that all the dependencies required to run PARCE are open source software, but some of them, such as Scwrl4, require academic licences. In such cases, it is recommended to install these packages following the developer's documentation to integrate their paths to the code. A list of all the required software, including the scoring functions, is provided in Table \ref{tbl:table1}.  All the steps associated to the mutation protocol, the sampling of the peptide-protein system and the application of scoring functions have been optimized in previous works of specific systems, supporting the choice of the default parameters. \citep{Ochoa2018,Soler2018,Gladich2015}. PARCE has an MIT licence that allows for the distribution of the code and its improvement through new functionalities.

\begin{table}[hbt!]
\centering
\caption{List of third-party tools and scoring functions required to run the PARCE.}
\label{tbl:table1}
      \begin{tabular}{ll}
        \hline
        Name & Version/Year \\
        \hline
        Gromacs & 5.1.4 \\
        Scwrl & 4.0  \\
        GromacsWrapper (Python3) & 0.8 \\
        BioPython (Python3) & 1.76 \\
        PDB2PQR & 2.1.1 \\
        Bluues & 2.0 \\
        BMF & 3.0 \\
        Pisa & 2011 \\
        Zrank & 2007 \\
        BACH & 6.0 \\
        DFIRE-GOAP & 2011 \\
        FireDock & 2007 \\
        Irad & 2011 \\
     \hline
   \end{tabular}
\end{table}

\subsection{Design of peptides bound to a protease}

As a tutorial, we use a protease in complex with a modelled peptide. The protease is obtained from the crystallized structure (PDB code 1ppg). This is a neutrophil elastase, which is a serine protease part of the chymotrypsin family \citep{an1988refined}. The enzyme is bound to a peptidomimetic inhibitor, so we modified its sequence to model a reported peptide-substrate available in the  MEROPS database \citep{Rawlings2018}. Specifically, the protease binding pockets were annotated according to its catalytic site. Then, the modified amino acids were replaced by natural amino acids found in the substrate sequence using Scwrl4. The peptide was completed to reach a final size of eight amino acids (AAPAAAPP), which is characteristic of these proteas
e peptide binders \citep{Schechter1967}. These missing amino acids were predicted with the Modeller software \citep{Marti-Renom2000}. The created complex was subjected to 100 ns of MD simulations using the same MD configuration as explained previously. We ran the PARCE protocol to perform 30 random mutation attempts, using six scoring functions with a consensus threshold of three. The evolution of the scores was monitored to check if the scores -on average- decreased as a function of the MC iteration step.

\begin{figure*}[hbt!]
\centering
\includegraphics[width=1.0\textwidth]{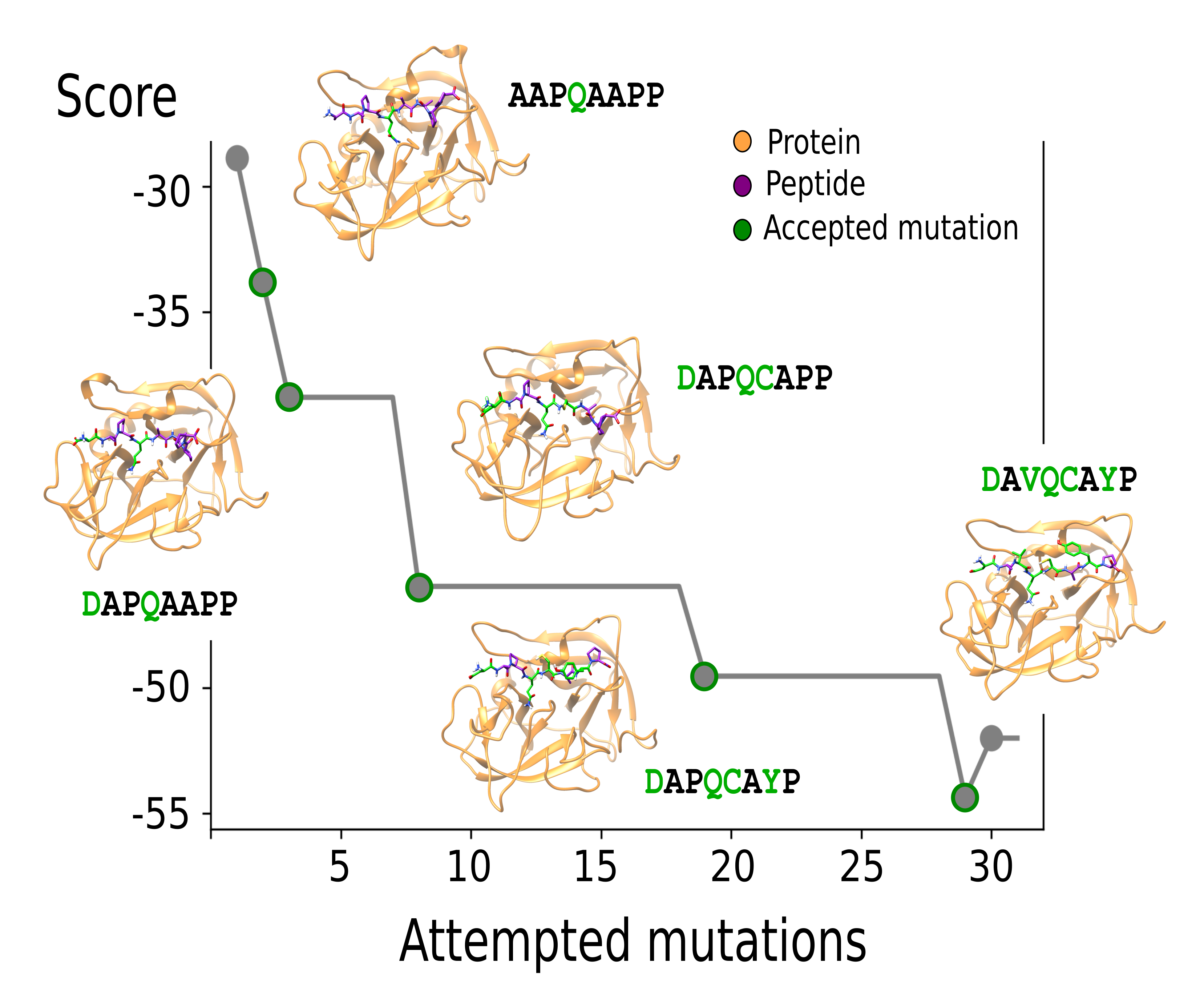}
\caption{Example of a PARCE run showing one (BMF-Bluues) of the 6 scoring functions with the protease-peptide system (see the Methods). The dots represent the mutations that were accepted, with the structure of the protein in orange, the peptide in purple, and the accepted mutations in green.}
\label{fig:Fig3}
\end{figure*}

The protocol was ran using 5 ns of conformational sampling per mutation for 30 attempts. For this specific run, a total of 6 mutations were accepted based on the consensus criterion explained in Methods (\textit{i.e.,} if three or more scores consider the mutation as favorable then it is accepted). During the PARCE run it is important to monitor if the scoring functions are -on average- lowering their values. An example is shown in Figure \ref{fig:Fig3} for the scoring function BMF-Bluues. The sequence of the peptide should evolve towards a sequence with potentially better affinity.

\begin{figure*}[hbt!]
\centering
\includegraphics[width=1.0\textwidth]{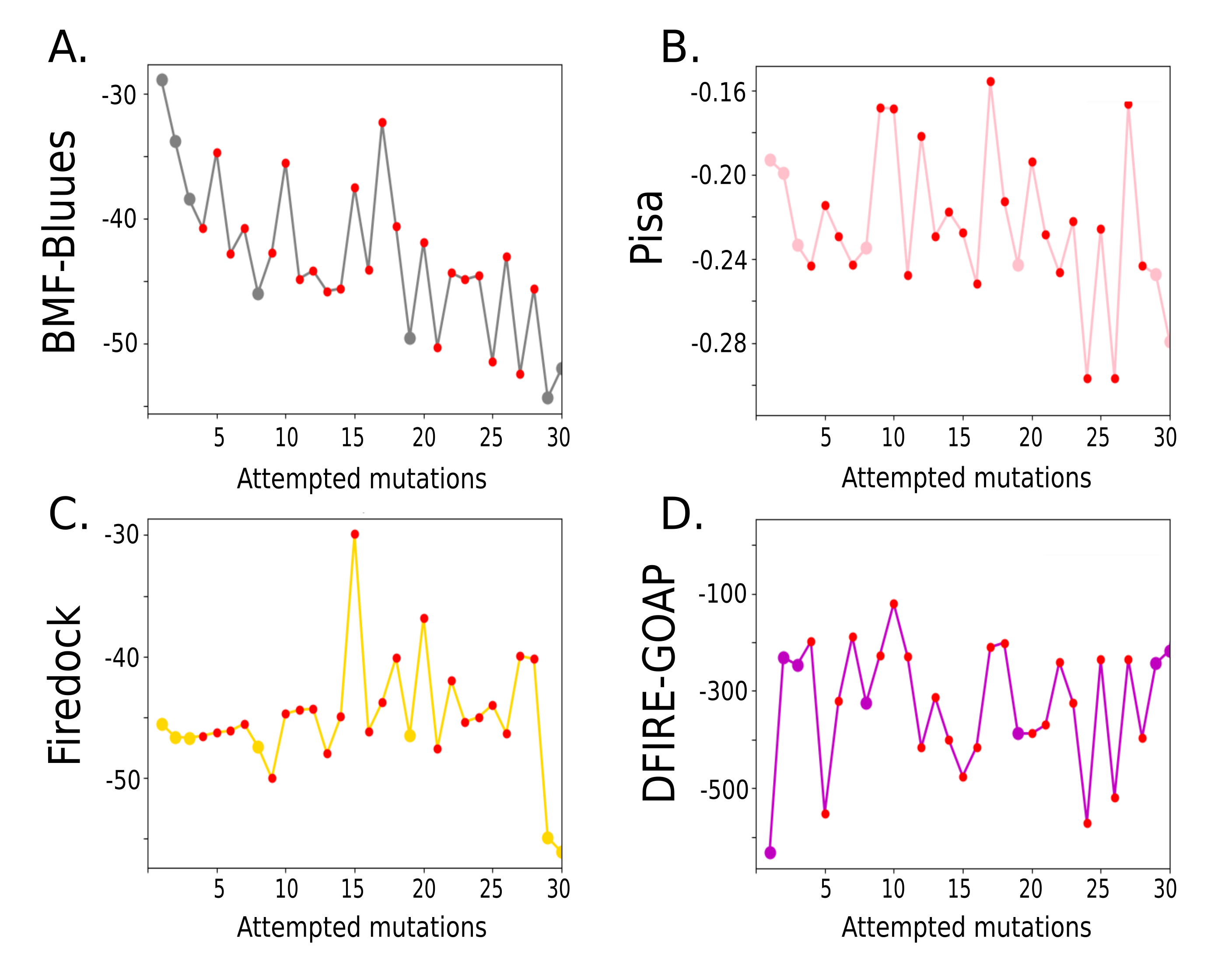}
\caption{Evolution of four of the six scoring functions used in PARCE for the example run with the protease-peptide system. The large and red dots correspond to the accepted and rejected mutations, respectively. The results are shown for (A) BMF-Bluues (same data as in Figure \ref{fig:Fig3}), (B) Pisa, (C) Firedock and (D) DFIRE-GOAP.}
\label{fig:Fig4}
\end{figure*}

Despite the ideal behavior of BMF-Bluues, some of the scoring functions have slightly different behaviours. An example of three other scoring functions behaviour is shown in Figure \ref{fig:Fig4}. For three of the four plotted functions, the scores get lower in the last steps. In the case of the DFIRE-GOAP combination (Figure \ref{fig:Fig4}D), the score fluctuates up and down without a defined trend. The consensus methodology ensures that on average most -but not all- of the scoring functions decrease. Therefore, if there is a poor-performing scoring function the method does not force it to minimize (\textit{e.g.}, DFIRE-GOAP in Figure \ref{fig:Fig4}). This is an advantage because some scoring functions might perform well for some particular systems but fail for others, and knowing beforehand which scoring functions are the best for a particular system is challenging. We note that the evolution is stochastic (due to the random mutations) therefore for different runs there could be different behaviors. However, if most of the scoring functions have an erratic performance, then the scoring function configuration should be re-evaluated. The user has the possibility to modify the selection of the scoring functions and the threshold to define if a mutation is accepted or not. Moreover, if the system has available binding data, it might be useful to benchmark the MD/scoring methodology with it. We applied the latest in the case of the Major Histocompatibility Complex class II (MHC-II), where we benchmarked a set of scoring functions and MD configurations to rank bound peptides in agreement with available experimental data \cite{Ochoa2019}.

We note that we have tested protocol extensively over other protein-peptide systems \cite{Guida2017,Soler2018}.

\subsection{Technical considerations}

Because the methodology is computationally expensive, the implementation can take weeks of calculations for running 50 or 100 mutation-attempts, depending on the system and the available computational resources. The PARCE computational cost is dominated by the MD simulation time (\textit{i.e.}, number of ns) multiplied by the number of mutation attempts. Other relevant factors are the system size, and the number of frames used to score the trajectories. We recommend to configure the protocol to have a sufficiently good MD sampling and a number of mutation-attempts sufficient to explore efficiently the peptide sequence space.

\section{Discussion}

The PARCE code is an open source software to design peptides or proteins capable of binding with higher affinity to a protein target. The method combines computational biophysics tools with bioinformatics, in order to achieve an equilibrium between accuracy and computational efficiency. One advantage is the possibility to obtain in an stochastic way peptide candidates, which can be different independent runs. This increases the pool of sequences available for further filtering and validation. This is an advantage against more deterministic or brute force alternatives. Moreover, since the code is open source, it is possible to improve it according to the research-project needs.

Most of the available protocols of peptide design have been built also with open source software, and some are available to users through web servers. This is the case of Peptiderive, which identifies the linear polypeptide segment that contributes most to the binding energy given a protein-protein interaction structure \citep{Sedan2016}.  Another server, PepComposer, requires only the target protein structure in order to retrieve a set of peptide-backbone scaffolds derived from monomeric proteins. Then, the peptides are optimized using a set of iterative mutations through controlled backbone movements \citep{Obarska-Kosinska2016}. Finally, other initiatives about the sampling and scoring peptides as binders are also available as web servers or code protocols that can be customized \citep{king2010structure,smith2008backrub,Raveh2010}.

Despite the efforts, the balance between the computational efficiency and biological accuracy is still a major challenge, motivating the development and validation of novel pipelines as the one proposed here. PARCE can accelerate the discovery of novel peptides as potential therapeutics, biomarkers or vaccine sub-units. The code is flexible. It allows adding protocols to modify different types of molecular targets, such as small molecules, contributing to the engineering of peptides and proteins for a broad spectrum of applications.

%\section{Declarations}

%\subsection{List of abbreviations}

%MD: Molecular Dynamics; PDB: Protein Data Bank; PME: Particle Mesh Ewald.

%\subsection{Competing Interests}

%The author(s) declare that they have no competing interests.

%\subsection{Funding}

%R.O. and P.C. were supported by Colciencias, University of Antioquia, Ruta N, Colombia, and the Max Planck Society, Germany. 

%\subsection{Author's Contributions}

%R.O designed and wrote the code, the documentation and the manuscript. M.A.S contributed with previous projects inspiring the creation of PARCE, reviewed the code and reviewed the manuscript. A.L provided the original idea, feedback of the implementation and reviewed the manuscript. P.C reviewed the code and wrote the manuscript.

\section{Acknowledgements}
R.O. and P.C. were supported by Colciencias, University of Antioquia, Ruta N, Colombia, and the Max Planck Society, Germany. The computations were performed in a local server with an NVIDIA Titan X GPU. P.C. gratefully acknowledges the support of NVIDIA Corporation for the donation of this GPU. Authors acknowledge Prof. Fogolari from University of Udine for providing the scoring functions Bluues and BMF.

\nocite{*}
\bibliographystyle{elsarticle-num}
\bibliography{biblio}

\appendix
\setcounter{table}{0}
\renewcommand*\thetable{\Alph{section}\arabic{table}}
\section{Supplementary tables}

\begin{table}[hbt!]
\caption{Parameters provided by the user in the configuration file.}
\label{tbl:suptable1}
      \begin{tabular}{p{3cm} p{11cm}}
        \hline
        Parameter & Explanation \\
        \hline
        folder & Name of the folder that has all the input and output files of the protocol \\
        mode & The design mode, which has three possible options: start (start the protocol from zero), restart (start from a particular iteration of a previous run) and *nothing* (just run without modifying existing files) \\
        peptide\_reference & The sequence of the peptide, or protein fragment that will be modified \\
        pdbID & Name of the structure that is used as input, which contains the protein, the peptide/protein and the solvent molecules \\
        chain & Chain id of the peptide/protein in the structural complex \\
        sim\_time & Time in nanoseconds that will be used to sample the complex after each mutation \\
        num\_mutations & Number of mutations that will be attempted \\ try\_mutations & Number of mutations tried after having minimization or equilibration problems \\
        residues\_mod & These are the specific positions of the residues that want to be modified. This depends on the peptide/protein length and the numbering in the PDB file \\
        md\_route & Path to the folder containing the input files, which are the files used during the previous MD sampling of the system \\
        md\_original & Name of the system file located in the folder containing the previous MD sampling \\
        score\_list & List of the scoring functions that will be used to calculate the consensus. Currently the package only has available the code for six of them: BACH, Pisa, ZRANK, IRAD, BMF-BLUUES and FireDock \\
        half\_flag & Flag that controls which part of the trajectory is used to obtain the average score. If 0, the full trajectory is used, if 1, only the last half \\
        threshold & Threshold used for the consensus. If the number of scoring functions in agreement are equal or greater than the threshold, then the mutation is accepted. \\
        scwrl\_path & Provide the path to Scwrl4 in case it is not installed in a PATH folder. \\
        gmxrc\_path & Provide the path to GMXRC for Gromacs.  \\

     \hline
   \end{tabular}
\end{table}

\begin{table}[hbt!]
\caption{Example of PARCE report using as input a peptide binder with sequence AAPFAAPP. The report includes the iteration number, the mutation, the acceptance, the average scores and the attempted peptide sequence.}
\label{tbl:suptable2}
\resizebox{\textwidth}{!}{
      \begin{tabular}{llllllllll}
        
    \hline
        Iteration & Mutation & Status & BACH & Pisa & Zrank & Irad & BMF-BLUUES & Firedock & Sequence \\
    \hline
        1 & AB4F & Accepted & -3.514 & -0.209 & -59.637 & -101.285 & -30.542 & -44.041 & AAPFAAPP \\
        2 & AB2G & Rejected & -4.705 & -0.198 & -56.245 & -101.161 & -34.482 & -37.784 & AGPFAAPP \\
        3 & PB3Q & Accepted & -4.128 & -0.273 & -63.155 & -108.444 & -34.122 & -46.402 & AAQFAAPP \\
        4 & PB8R & Rejected & -3.568 & -0.237 & -56.869 & -100.154 & -32.223 & -37.734 & AAQFAAPR \\
        5 & AB1Q & Accepted & -3.463 & -0.245 & -65.987 & -116.340 & -35.730 & -47.378 & QAQFAAPP \\
     \hline
   \end{tabular}}
\end{table}

\begin{table}[hbt!]
\caption{Folders with the output files generated by PARCE.}
\label{tbl:suptable3}
      \begin{tabular}{p{3cm} p{10cm}}
        \hline
        Folder & Explanation \\
        \hline
    binder & Store the peptide/protein structure after each mutation attempt \\
    target & Store the target structure after each mutation attempt \\
    complexP & Store the target-peptide/protein structure after each mutation attempt \\
    solvent & Store the solvent box after each mutation attempt \\
    system & Store the complete target-peptide/protein-solvent complex after each mutation attempt \\
    trajectory & Store the MD trajectory of the previous mutations \\
    score\_trajectory & Store the average scores for each snapshot from the trajectories. The file is split into four columns. The first column is the score of the complex. The second and third are the scores for the receptor and peptide alone. The fourth column is the total score after doing the difference between the complex and each component \\
    log\_npt & Store the log file from each npt run to verify possible errors \\
    log\_nvt & Store the log file from each nvt run to verify possible errors \\
     \hline
   \end{tabular}
\end{table}


\end{document}